\documentclass[10pt,pra,aps,amssymb,amsmath,tightenlines,twocolumn]{revtex4} 

\usepackage{graphicx,epsfig,xcolor}
\usepackage[T1]{fontenc}
\usepackage[latin1]{inputenc}
\usepackage[english]{babel}
\usepackage{amsmath, bbm, nccmath, MnSymbol}
\usepackage{mathtools}
\usepackage{amsfonts}
\usepackage{amssymb}
\usepackage{bm}
\usepackage{braket}
\usepackage{dsfont}
\usepackage[normalem]{ulem} 
\usepackage{float}
\usepackage{color}
\definecolor{ikb}{rgb}{0, 0.184,0.655}

\usepackage[labelformat=parens]{subfig}
\usepackage[justification=raggedright, font=footnotesize]{caption}
\usepackage[colorlinks=true, linkcolor=ikb, urlcolor=ikb, citecolor=ikb,
            anchorcolor=ikb]{hyperref}

\newcommand{\cP}{\ensuremath{\mathcal{P}}}

\newcommand{\cH}{\ensuremath{\mathcal{H}}}

\newcommand{\cT}{\ensuremath{\mathcal{T}}}
\newcommand{\cC}{\ensuremath{\mathcal{C}}}

\newcommand{\cPT}{\ensuremath{\mathcal{PT}}}
\newcommand{\cCPT}{\ensuremath{\mathcal{CPT}}}

\begin{document}

\title{Three perspectives on entropy dynamics in a non-Hermitian two-state system}
\author{Alexander Felski$^a$\footnote{Corresponding author}}\email{alexander.felski@mpl.mpg.de}
\author{Alireza Beygi$^b$}\email{alirezabeygi389@gmail.com}
\author{Christos Karapoulitidis$^c$}\email{chris.karapoulitidis@gmail.com}
\author{S.~P.~Klevansky$^c$}\email{spk@physik.uni-heidelberg.de}

\affiliation{$^a$Max Planck Institute for the Science of Light, 91058 Erlangen, Germany\\
$^b$Institute of Computer Science, Goethe University Frankfurt, 60325 Frankfurt am Main, Germany\\
$^c$Institut f\"ur Theoretische Physik, Universit\"at Heidelberg, 69120
Heidelberg, Germany}

\begin{abstract}
A comparative study of entropy dynamics as an indicator of physical behavior in an open two-state system with balanced gain and loss is presented. 
To begin with, we illustrate the phase portrait of this non-Hermitian model on the Bloch sphere, 
elucidating the changes in behavior as one moves across the phase transition boundary, as well as the emergent feature of unidirectional state evolution in the spontaneously broken $\cPT$-symmetry regime.
This is followed by an examination of the purity and entropy dynamics. Here we
distinguish the perspective taken in utilizing the conventional framework of Hermitian-adjoint states from 
an approach that is based on biorthogonal-adjoint states and a third case based on an isospectral mapping.
In this it is demonstrated that their differences are rooted in the treatment of the environmental coupling mode. 
For unbroken $\cPT$ symmetry of the system, a notable characteristic feature of the perspective taken is the presence or absence of purity oscillations, with an associated entropy revival.    
The description of the system is then continued from its $\cPT$-symmetric pseudo-Hermitian phase into the 
regime of spontaneously broken symmetry, in the latter two approaches through a non-analytic operator-based continuation, yielding a Lindblad master equation based on the $\cPT$ charge operator $\cC$. This phase transition indicates a general connection between the pseudo-Hermitian closed-system and the Lindbladian open-system formalism through a spontaneous breakdown of the underlying physical reflection symmetry. 
\end{abstract}
\maketitle

\section{Introduction}
\label{s1}

Traditional quantum mechanics describes Hermitian systems. Yet many physical models have, at times effective, non-Hermitian Hamiltonians.

Following the discovery that parity-time ($\cPT$) symmetry in non-Hermitian quantum-mechanical systems can lead to a real energy spectrum \cite{BB98}, the theory of $\cPT$-symmetric and pseudo-Hermitian systems has been developed as an intermediary between conventional Hermitian and  non-Hermitian cases \cite{B2007, B2019, B-2014, M-2010, EMKMRC-2018}. Such theories can admit both phases of spectral reality 
(the symmetric or unbroken regime) 
and regions in which eigenvalues arise in complex conjugate pairs 
(the phase of spontaneously broken symmetry).
The mathematical framework of pseudo-Hermiticity establishes a general connection between real energy spectra and Hermitian behavior in non-trivial Hilbert spaces, while $\cPT$ theory proposes a physical foundation of quantum mechanics in the concept of an underlying time-reversing reflection symmetry. 

To describe the dynamics of non-Hermitian quantum systems, three interrelated approaches are commonly used: 
(a) treating them as open subsystems within the standard Hermitian framework; 
(b) treating them as closed systems with a non-trivial state space by introducing a set of biorthogonal adjoint eigenstates; 
(c) utilizing a similarity transformation to render the underlying Hamiltonian isospectrally Hermitian. 
In all approaches, addressing the spontaneously symmetry-breaking phase transition and the description of the broken symmetry regime give rise to open questions.
In particular, the properties of the energy spectra of $\cPT$-symmetric systems are by now well understood;
the three formalisms display the property of \emph{spectral equivalence}.
But the concept of entropy is the subject of ongoing discussion \cite{FF2019, KAU2017, SZ2016, JM2010, MGK2024, CK2023, MCM2023}, since no such equivalence property exists between the approaches.

We present a detailed investigation of the ubiquitous two-state system, modified through the inclusion of balanced gain and loss terms, with the purpose of illustrating the physical differences and limitations of each of the three approaches. 
The focus here lies in the discussion of the phase dynamics of the eigenstates, and the resulting dynamics of the relevant density matrices and entropies that one can define.
We also discuss, in particular, the phenomena of purity oscillations and asymptotic purification, which arise as a result of the environmental coupling mode in the system and the perspective taken in incorporating this coupling through the lens of the approach used.
Due to the crucial role of environmental interactions in experimental settings, such phenomena have attracted interest in the context of quantum information processing \cite{WZZ2021, PGUWZ2003, WTHKGW2011, DMP2021} as well as transport processes in quantum systems \cite{SRDCW-2019, DHCW-2016, KMCW2013, LHM2019}. Recent applications include non-Hermitian metrology \cite{XY2023, XY2024} and quantum speed limits for non-Hermitian quantum systems \cite{FI2021, DT2024}.
The ubiquity of the two-state system and its toy-model character here signal potential implications of a generic understanding of these effects for a wide range of applications in the future.

Our study explicitly addresses the continuation of all of these concepts into the regime of spontaneously broken $\cPT$ symmetry
through the introduction of a novel operator-based continuation of the $\cPT$ charge operator $\cC$ for this region,
which allows for a direct comparison of the behavior resulting within all three frameworks (a) to (c) in both the phases of unbroken and spontaneously broken symmetry.
It
demonstrates the co-limiting nature of pseudo-Hermiticity and the construction of a probabilistic -- that is in this case a positive definite biorthogonal -- formulation in 
the broken regime, 
and argues for the connection between open and closed system dynamics established by an underlying spontaneous breakdown of the time-reversing reflection symmetry.

This paper is organized as follows:
In Secs.~\ref{s2} and \ref{s3} the two-state system with balanced gain and loss is introduced briefly, and its phase dynamics are illustrated. Sections~\ref{s4}, \ref{s5} and \ref{s6} then discuss the dynamics of the density matrix, the purity and the entropy, for the model based on the Hermitian-adjoint, biorthogonal and isospectral-mapping formalism respectively. We conclude and summarize the main findings in Sec.~\ref{s7}.

\section{Non-Hermitian two-state Hamiltonian}
\label{s2}

We begin with the two-state Hamiltonian,
\begin{equation}
\label{s2e1}
	\cH =  \begin{pmatrix}
		r \,\mathrm{e}^{i \theta} & d \\
		d & r\,\mathrm{e}^{-i\theta}
	\end{pmatrix}
\, ,
\end{equation}
describing an open system with balanced gain and loss rates in which the states are coupled to one another with the direct coupling strength $d$, and to the environment through the imaginary contributions $\pm i r \sin\theta$ \cite{B2007}. This system is non-Hermitian but symmetric under combined parity reflection $\cP$ and time reversal $\cT$, where
\begin{equation}
\label{s2e2}
    {\cP} = 
    \begin{pmatrix}
        0 & 1 \\
        1 & 0
    \end{pmatrix}, \qquad \text{so that} \quad \cP^2=\mathbbm{1};
\end{equation}
the operator $\cT$ performs complex conjugation, inverting gain and loss, and therefore
$\cT^2=\mathbbm{1}$ as well. 

The eigenvalues of the Hamiltonian $\cH$ have the form
\begin{equation}
\label{s2e3}
    E_{\pm} = r\cos \theta \pm(d^2-r^2 \sin^2\theta)^{1/2}
\end{equation}
and exceptional points arise as square-root branch points when $d^2= r^2 \sin^2\theta$, separating parametric regions of real and complex energies that characterize regimes of unbroken and spontaneously broken $\cPT$ symmetry.
The Schr\"odinger equation
\begin{equation}
\label{s2e4}
    i\partial_t\vert\psi\rangle = \cH\vert\psi\rangle
\end{equation}
governs the evolution of the (right-hand) quantum states.

\section{Bloch Sphere and Phase portrait}
\label{s3}

The dynamics of any pure state $\ket{\varphi} = (\varphi_1, \varphi_2)^T$ can be expressed in terms of the polarization components $z= \varphi_1/\varphi_2$ or $w=z^{-1}$, normalized as
\begin{equation}
\label{s3e1}
    \ket{\varphi}
    \mapsto
    \frac{1}{\sqrt{1+z z^*}} \begin{pmatrix}
         z \\
         1
 \end{pmatrix} ,
    \quad 
	\ket{\varphi}
    \mapsto
    \frac{1}{\sqrt{1+w w^*}} \begin{pmatrix}
        1 \\
        w
 \end{pmatrix} ,
\end{equation}
with $z^*$ being the complex conjugate of $z$.
The real and imaginary contributions of these variables span phase portraits in the complex plane with origin 
\begin{figure}[t]
\centering
\includegraphics[width=0.48\textwidth]
{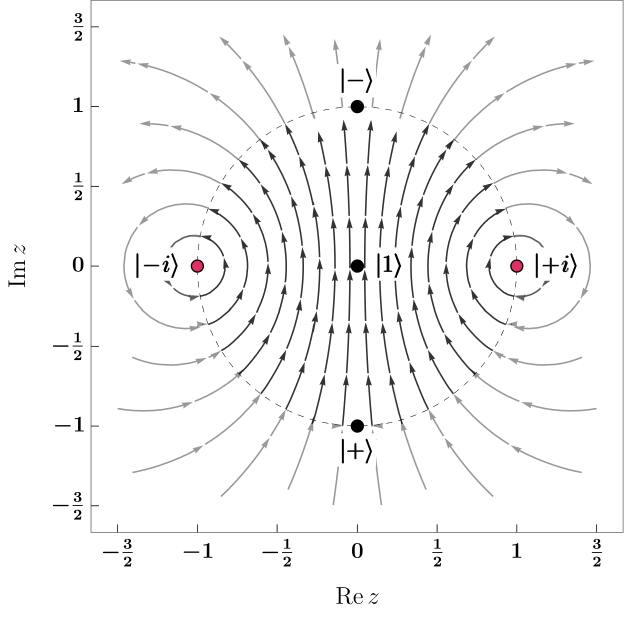}
\caption{
\label{f1}
Phase portrait for the Hermitian limit $\theta=0$, spanned by the real and imaginary parts of the polarization $z$. The positions of common superposition states are shown as black dots for reference.
Stationary points (centers) are indicated as red dots.
The dynamics within the unit disk correspond to the behavior on the lower Bloch hemisphere.
}
\end{figure}
\begin{equation}
\label{s3e2}
    \ket{1} = (0,1)^T
    \qquad \text{and} \qquad
    \ket{0} = (1,0)^T
\end{equation}
respectively.
From the equations of motion for $z$ and $w$,
\begin{align}
\label{s3e3}
	\partial_t z =& \, i \, (\, {d}z^2 - 2ir\sin \theta\, z - d \,)  ,\\
    \partial_t w =& \, i \, (\, d w^2 + 2ir\sin \theta\, w - {d} \,),
\end{align}
the stationary points of the state evolution are identified as being
\begin{align}
\label{s3e4}
	z_c =&\, \tfrac{2}{d}\, [ir\sin\theta \pm (d^2-r^2\sin^2\theta)^{1/2}\,] ,\\
    w_c =&\, \tfrac{2}{d}\, [-ir\sin\theta \pm (d^2-r^2\sin^2\theta)^{1/2}\,].
\end{align}
An example of a phase portrait is shown in Fig.~\ref{f1} for the Hermitian limiting case of an isolated system when $\theta=0$. 
Here the critical points are given by the states 
\begin{equation}
\label{s3e5}
    \ket{+i} = \tfrac{1}{\sqrt{2}} (\ket{0} + i \ket{1} ) 
\quad \text{and} \quad
    \ket{-i} = \tfrac{1}{\sqrt{2}} ( \ket{0} -i \ket{1} ) ,
\end{equation}
which are centers of the pure-state dynamics indicated as red dots in the figure.

\begin{figure}[H]
\centering
\subfloat[\centering In the unbroken $\cPT$ symmetry regime. Stationary points (red) are centers.]{
\includegraphics[width=0.34\textwidth]
{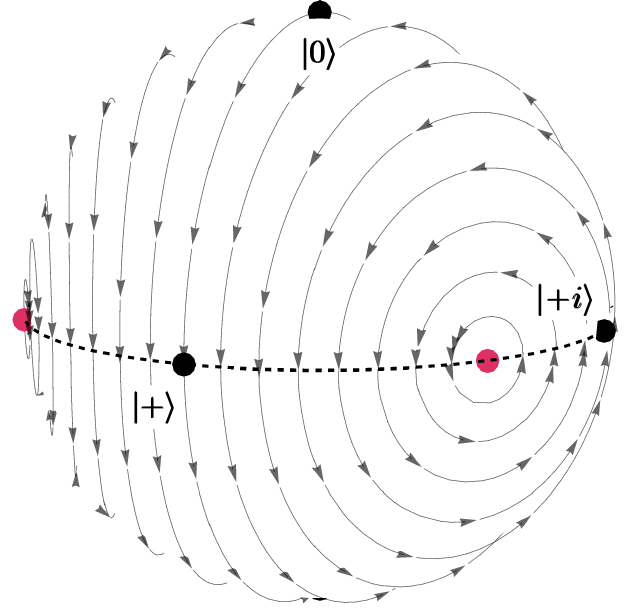}
\label{f2a}
}
\\[20pt]
\subfloat[\centering At the exceptional point. Stationary points (red) combine to a point dipole.]{
\includegraphics[width=0.34\textwidth]
{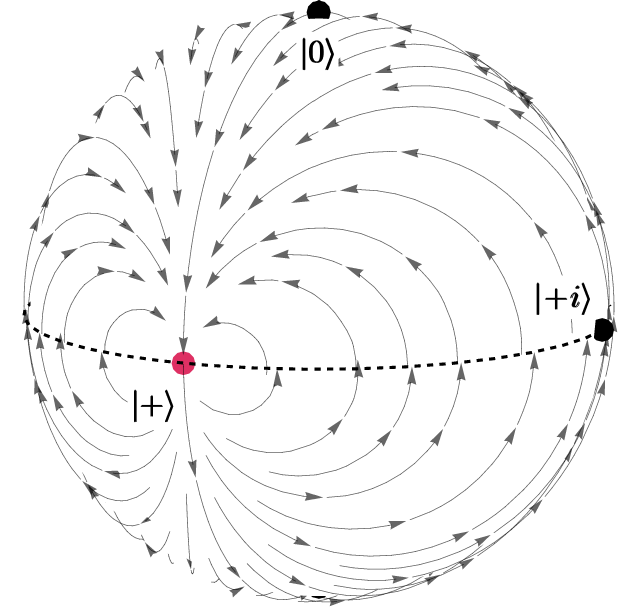}
\label{f2b}
}
\\[20pt]
\subfloat[\centering In the spontaneously broken $\cPT$ symmetry regime. Stationary points (red) are a source/sink.]{
\includegraphics[width=0.34\textwidth]
{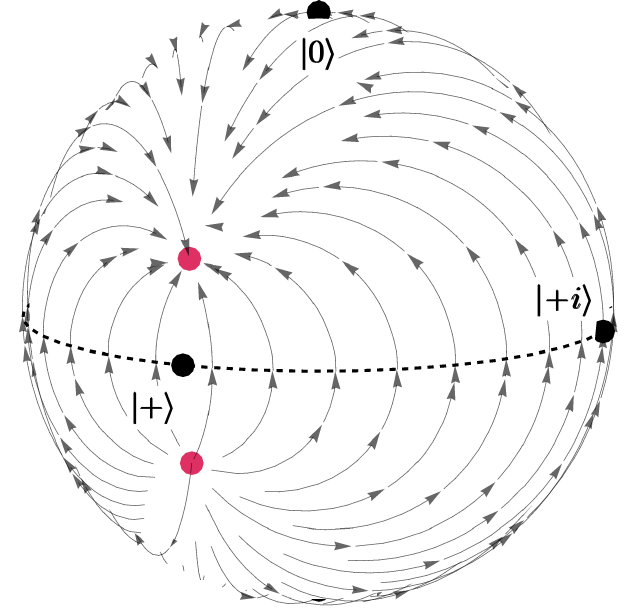}
\label{f2c}
}
\caption{
\label{f2}
(a) Phase dynamics on the Bloch sphere for the non- Hermitian model in the unbroken phase. ($r=1$, $d=1$, $\theta= \pi/5$)\\
(b) Phase dynamics at the phase transition. ($r=1$, $d=1$, $\theta= \pi/2$)\\
(c) Phase dynamics in the broken phase. ($r=1$, $d=0.95$, $\theta= \pi/2$)
\vspace*{-1cm}
}
\end{figure}

Changes of the phase dynamics due to the gain and loss terms can be visualized alternatively 
on the Bloch sphere. We choose $r=1$ for simplicity and show three examples illustrating the broken and unbroken symmetry regimes, as well as the exceptional-point case that separates them, in Fig.~\ref{f2}. 

In the unbroken-symmetry phase of the non-Hermitian model, $\theta \not\in \{0,\pi\}$, shown in Fig.~\ref{f2a}, the critical points remain centers which, under variation of the parameters $d$ and $\theta$, move along the equator of the Bloch sphere toward either 
\begin{align}
\label{s3e6}
    \ket{+} =& \tfrac{1}{\sqrt{2}} (\ket{0} + \ket{1}) ,
\quad \text{for} \quad \theta \in (0, \pi),
\quad \text{or}\\[3pt]
    \ket{-} =& \tfrac{1}{\sqrt{2}} (\ket{0} - \ket{1}) ,
    \quad \text{for} \quad \theta \in (\pi, 2\pi).
\end{align}

When the exceptional point is reached, that is  $d^2= r^2\sin^2\theta$, the two critical points coalesce and form a point dipole at either $\ket{+}$  or $\ket{-}$, the former case being illustrated in Fig.~\ref{f2b}.

Upon entering the broken symmetry regime, the critical points part again, moving into the upper and lower hemispheres toward the states $\ket{0}$ and $\ket{1}$, as shown in Fig.~\ref{f2c}. 
They now form a source and a sink of the state evolution \cite{LJS2021}.

This behavior can be understood intuitively:
In the unbroken symmetry regime, the direct coupling $d$ between the 
states $\ket{0}$ and $\ket{1}$   
is strong enough to balance out any gain and loss dynamics. 
The interaction with the environment reduces to another coupling mode of the two states in the system.
Any loss of density from one state to the environment is transferred through and regained from it by the other state in equal measure;
the environmental coupling mode is 
\emph{unidirectional}, but 
\emph{transitive}  only. 
In this regime, a perspective that encompasses both coupling channels then corresponds to a different  two-state model without gain or loss but with an effective direct coupling strength. This correspondence is the foundation of the isospectral mapping perspective that is discussed later in Sec.~\ref{s6}.
As in a closed coupled system, the state evolution displays a periodic behavior, reflected in the closed orbits of the phase dynamics. 

But balanced gain and loss \emph{rates} do not necessarily constitute a balanced gain and loss of density. 
In the spontaneously broken symmetry regime, the direct coupling $d$ is too small to balance the gain and loss dynamics anymore.
This results in an accumulation within the 
excited state over time while the ground state
drains (for $\theta \in (0,\pi)$), reflected also in the appearance of a source and sink of the state evolution.
The gain in the excited state can no longer be sourced from an equal loss from a drained ground state, so that the environment no longer merely transfers -- it feeds an overall gain of the density 
for the two-state system; the environmental coupling mode thus becomes \emph{generative} in the broken regime
and the overall dynamics become unidirectional \cite{REK2011, LFW2017}. 

Portraying the effects of the non-Hermiticity in the open 
\hfill two-state \hfill system \hfill through \hfill the \hfill phase \hfill dynamics \hfill on 
\begin{figure*}[t]
\centering
\includegraphics[width=0.95\textwidth]
{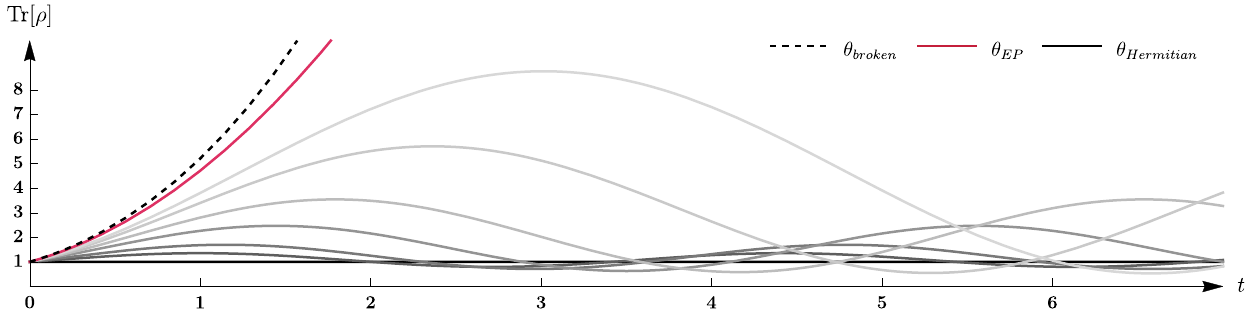}
\caption{
\label{f3}
Time evolution of $\mathrm{Tr}[\rho]$. Illustrative example with $r=1$, $d= 0.95$ and angles $\theta \in [0,\pi/2]$; the exceptional point lies at $\theta_{EP} \approx 1.253 $ (red). (Initial state $\ket{0}\!\bra{0}$.)
}
\end{figure*}
\break\noindent
the Bloch sphere demonstrates another essential property:
The eigenstate solutions of the time-independent 
Schr\"odinger equation, corresponding  to  the  stationary 
points of the phase dynamics, generally no longer appear as antipodal points when gain and loss rates are included. Thus, the eigenstates of the non-Hermitian theory do not form an orthogonal pair in the standard formalism.
Even though they are stationary solutions, their transition amplitudes do not vanish.

We illustrate these points in the following sections.

\section{Hermitian-adjoint formalism}
\label{s4}

Standard procedure in the conventional formalism for Hermitian quantum systems introduces the Hermitian adjoint (left-hand) states, which satisfy
\begin{equation}
\label{s4e1}
	-i \partial_t \bra{\psi} =  \bra{\psi} \cH^\dagger
\, .
\end{equation}

In the previous section, the normalization of the pure (right-hand) states $\ket{\varphi}$ in (\ref{s3e1}) implicitly made use of this, even though the formalism is applied to a non-Hermitian system;
these states satisfy $\braket{\varphi \vert \varphi} = \lvert \varphi_1 \rvert^2+\lvert \varphi_2 \rvert^2 = 1$.

This \emph{normalizes} the population of the two-state system such that the perspective taken is not one encompassing the two interacting states and the environment in which they are embedded -- it rather focuses on just the relative dynamics of the two states instead, in the following referred to as the subsystem.

\subsection*{Relative Dynamics of the Subsystem}

Making use of the (left-hand) states in (\ref{s4e1}) the evolution of the model can be reexpressed 
through the non-Hermitian equivalent of the Liouville-von Neumann equation \cite{SS2006, SZ2013}
\begin{equation}
\label{s4e2}
    \partial_t \rho = -i (\cH \rho - \rho \cH^\dagger) 
    = -i [\cH_h, \rho] + \{\cH_{ah}, \rho \} 
\end{equation}
in terms of the Hermitian and anti-Hermitian contributions of the Hamiltonian, $\cH_h$ and $i\cH_{ah}$ respectively, with $\cH = \cH_h + i \cH_{ah}$.
The general density matrix within this Hermitian-adjoint formalism is represented by
\begin{equation}
\label{s4e3} 
    \rho = \sum_n p_n \ket{\varphi_n}\!\!\bra{\varphi_n} , 
\end{equation}
chosen such that initially $\sum_n p_n = 1$.
We obtain the overall dynamics of the two-state subsystem through the time evolution of the trace of the density matrix
$\mathrm{Tr[\rho]}$, which notably is \emph{not conserved} due to the anticommutator with the anti-Hermitian portion $\cH_{ah}$ of the Hamiltonian,
\begin{equation}
\label{s4e4}
    \partial_t \mathrm{Tr}[\rho] = 
    2 \,\mathrm{Tr}[ \cH_{ah}\, \rho ] = 
    2\braket{\cH_ {ah}} 
\, .
\end{equation}

Figure~\ref{f3} illustrates the time evolution of this trace. 
In the unbroken symmetry regime an \emph{oscillatory} behavior is found. 
This arises as a modification to the generic quantum (Rabi) oscillations in a Hermitian two-state system with direct coupling. 
Here the gain and loss rates $\pm ir\sin\theta$ introduce an imaginary detuning which \emph{reduces} the generalized transition frequency $\Omega = (d^2-r^2\sin^2\theta)^{1/2}$ due to the periodic transfer between the two states through the environmental coupling mode. 
When focusing on the subsystem dynamics, this exchange between the two-state subsystem and the environment becomes apparent as a periodic deviation from the constant trace of the density matrix in a Hermitian model.
Thus, values $\mathrm{Tr[\rho]} > 1$ can be reached, while in general $\mathrm{Tr[\rho]} \geq 0$ still.
We emphasize that this is a feature of the open-subsystem perspective taken in the Hermitian adjoint formalism.
Nonetheless, the oscillations mark a \emph{weak conservation} of density due to the realization of a transitive environmental coupling mode that preserves a static average.
In the Hermitian limit $\theta \rightarrow 0 \lor \pi$ the amplitude of these oscillations vanishes, and we recover the conventional static unit trace, shown as a solid black line in Fig.~\ref{f3}.

\begin{figure*}[t]
\centering
\subfloat[\centering purity]{
\includegraphics[width=0.47\textwidth]
{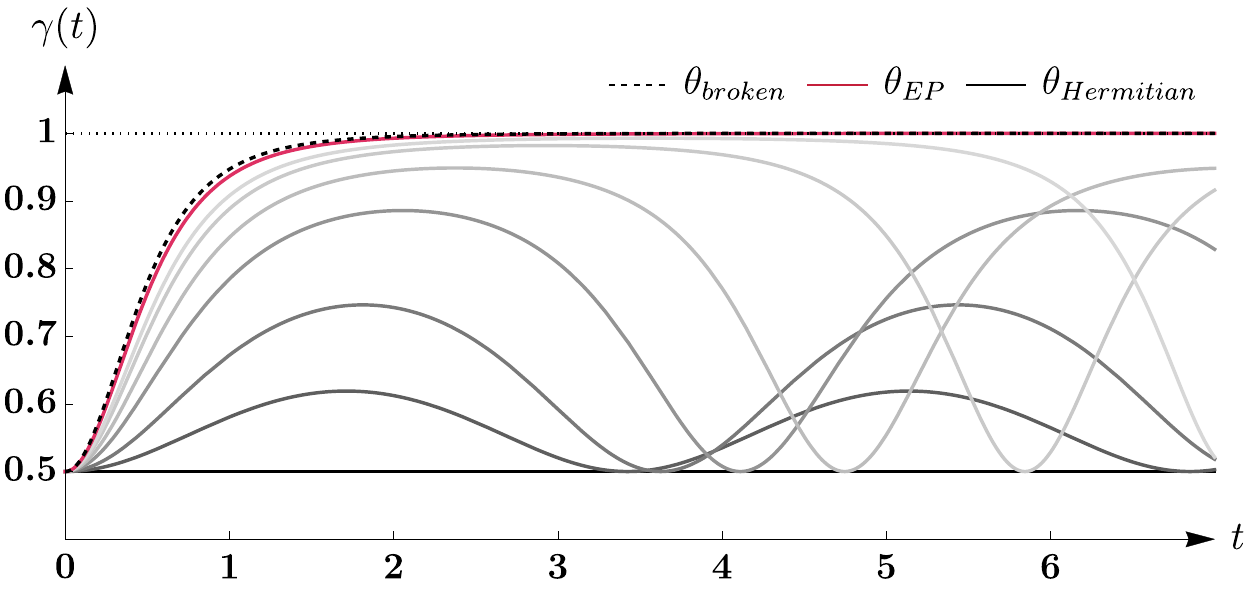}
\label{f4a}
}
\subfloat[\centering entropy]{
\includegraphics[width=0.47\textwidth]
{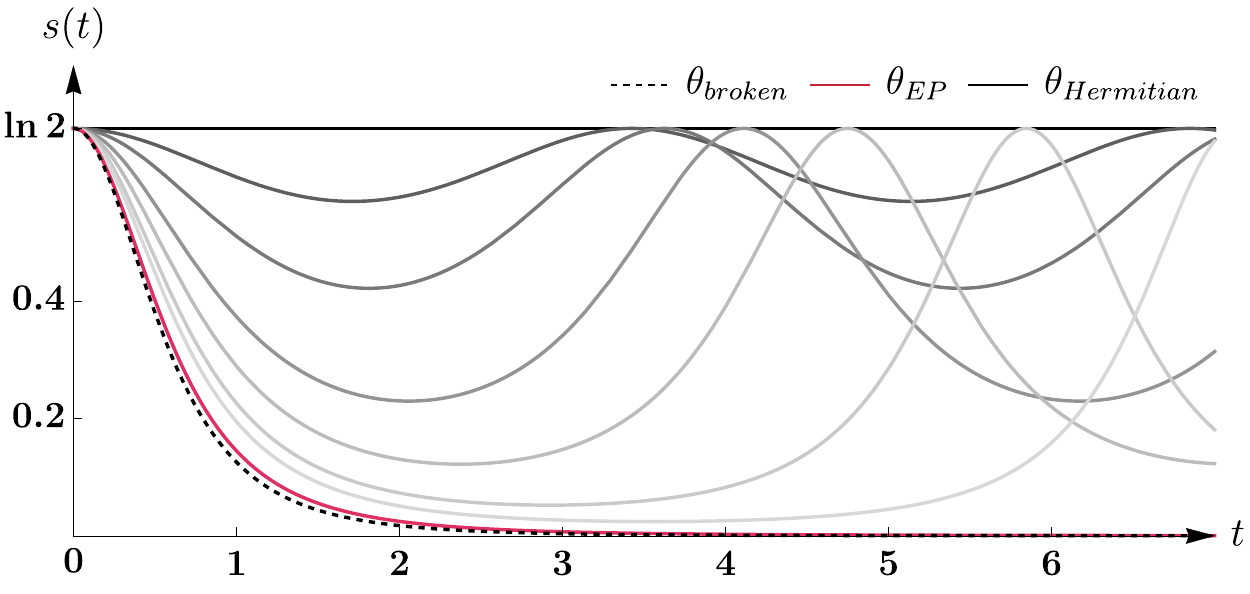}
\label{f4b}
}
\caption{
\label{f4}
Time evolution of the purity $\gamma(t)$ and the entropy $s(t)$.  Illustrative example with $r=1$, $d= 0.95$ and angles $\theta \in [0,\pi/2]$; the exceptional point lies at $\theta_{EP} \approx 1.253 $ (red). (Maximally mixed initial state, 
representative of generic mixed initial state behavior.)
}
\end{figure*}
Upon undergoing the spontaneously $\cPT$ symmetry breaking phase transition at the exceptional point, shown in red in Fig.~\ref{f3}, 
the generalized transition frequency $\Omega$ vanishes and the period of oscillation becomes infinite.
The 
behavior of $\mathrm{Tr[\rho]}$ changes into an 
unbounded (polynomial)
asymptotic growth, 
which becomes exponential throughout the broken symmetry phase,
illustrated as a dashed black line. It reflects the generation of density within the subsystem sourced from the environment.
This behavior is qualitatively independent of the initial states of the evolution.

Despite the time dependence of $\mathrm{Tr[\rho]}$, the rank of $\rho$ remains unchanged at finite times.
Since a density matrix of $\mathrm{rank}(\rho) = 1$ represents pure quantum states,  
a pure initial state remains pure under time evolution in this non-Hermitian model \cite{SZ2013}. 
Moreover, for the established relation that characterizes the purity of a \emph{normalized} quantum state, 
\begin{equation}
\label{s4e5}
    \gamma(t) = 
    \mathrm{Tr}[\tilde{\rho}^2(t)] = 1  
\quad  \text{iff} \,\, \rho \,\, \text{is pure},
\end{equation}
it becomes essential to utilize the normalized density matrix $\tilde{\rho}(t) = \rho(t)/\mathrm{Tr}[\rho(t)]$ in the context of the non-Hermitian system \cite{SZ2013, SZ2016}.
Accordingly, the representation of the state evolution on the Bloch sphere in the previous Sec.~\ref{s3} was based on normalized quantum states.

The behavior of $\gamma(t)$ is exemplified in Fig.~\ref{f4a} for a maximally mixed initial state 
$\rho(0) = \tfrac{1}{2} (\ket{0}\!\bra{0} + \ket{1}\!\bra{1})$ with initial purity $\gamma_{min}= \tfrac{1}{2}$. In the unbroken symmetry phase, the inclusion of the environmental interactions of the system causes an increase to values $\gamma_{min}<\gamma(t)<1$ with a periodic return to the initial value \cite{SRDCW-2019,DHCW-2016}. 

Since the normalization of states stresses the relative dynamics of the two-state subsystem, the density matrix (\ref{s4e3}) within this formalism takes the effective role of a reduced density matrix that specifies a subsystem state. 
A mixed subsystem state can thus be understood as a reduced state of a pure total state in a larger system that includes the environment.
As such, the mixedness of 
subsystem
states, reflected in their purity (\ref{s4e5}), indicates an entanglement of the subsystem with the 
(external) 
environment in which it is embedded--and with which it interacts. 
The periodic deviation from and revival to $\gamma_{min}$ then reflects the changing entanglement structure of the open system, which in the unbroken symmetry regime undergoes quantum oscillations between the two-state subsystem and the environment, as illustrated earlier in the behavior of $\mathrm{Tr}[\rho]$.

Upon transitioning into the broken symmetry phase, however, we find an \emph{asymptotic purification} of the initial state instead \cite{PGUWZ2003, WTHKGW2011, CM2021-II}, with $\gamma(t)$ increasing toward $1$. 
This reflects the accumulation within the excited state (for $\theta \in (0,\pi)$) in the presence of a generative environmental coupling.

The behavior of the entanglement structure of the system is apparent furthermore in the von Neumann entropy which, when based upon the reduced density matrix (\ref{s4e3}), takes the effective role of an entanglement entropy within the Hermitian adjoint formalism \cite{HHHH-2009}.

In correlation with the definition of the purity (\ref{s4e5}), the use of normalized quantum states appears pertinent, so that the von Neumann entropy is given by
\begin{equation}
\label{s4e6}
    s(t) = 
    \langle\, \ln[1/\tilde{\rho}(t)] \,\rangle = 
    -\mathrm{Tr}[\tilde{\rho}(t) \ln( \tilde{\rho}(t) )] 
\vspace{0.2cm}
\end{equation}
for the non-Hermitian model \cite{SZ2016, ZL2022}.
This vanishes in the case of pure states and reaches an upper bound of $\ln 2$ for a maximally mixed state, as for a Hermitian model. 
The time evolution starting from this latter case is illustrated in Fig.~\ref{f4b}. 
A periodic revival of the initial entropy value marks the unbroken symmetry regime, while the purification of the initial state in the broken symmetry regime is mirrored in an asymptotically vanishing entropy of entanglement between the subsystem and the external environment, illustrated in red.
Once again, this is representative of the general behavior.

Both the von Neumann entropy of entanglement and the purity can be regarded as special cases of the R\'enyi entropy, 
\begin{equation}
\label{}
    s_\alpha(t) = \mfrac{1}{1-\alpha}\ln ( \, \mathrm{Tr} [\tilde{\rho}^\alpha (t)] \,) ,
\end{equation}
for $\alpha=1$ and $\alpha=2$ respectively.
This establishes a connection to other commonly considered entropy measures, as well as the full entanglement spectrum \cite{LH-2008}.

\subsection*{Orthogonality}

The standard procedure in the conventional formalism for Hermitian quantum systems does, however, not transfer to the application within a non-Hermitian context without a caveat.
While the focus on the relative dynamics of the subsystem within the Hermitian adjoint formalism presents a worthwhile perspective that clearly delineates the influence of an environmental coupling from that of the direct coupling  mode, 
the use of Hermitian adjoint (left-hand) states does not facilitate a consistent use of standard projection methods and perturbation theory \cite{SW-1972, M1972, B-2014}.
For in addition to the existence of separate symmetry regimes with distinct dynamical behaviors, the open two-state system with balanced gain and loss displays another notable difference between non-Hermitian and Hermitian theories: 
the eigenstates associated
with the energy eigenvalues (\ref{s2e3}) of the Hamiltonian (\ref{s2e1}), \linebreak
\vspace*{-0.6cm}
\begin{equation}
\vspace*{-0.1cm}
\label{s4e7}
	\ket{\varphi_\pm} = \, \bigl( i r \, \sin\theta\, 
    \pm (d^2-r^2\sin^2\theta)^{1/2} ,\, d \bigr)^T
    \bigr/ \mathcal{N}_\pm  
\, ,
\end{equation}
which are normalized within the Hermitian-adjoint formalism by
\begin{widetext}

\begin{align}
\label{s4e7-2}
\begin{split}
    \mathcal{N}_\pm = 
    \sqrt{2}\, d \quad
&\quad \text{if} \,\,\,\,
    d^2-r^2\sin^2\theta >0 \quad  (\text{unbroken symmetry}) \\
    \mathcal{N}\pm = 
    \bigl(-2 i r \sin\theta \, [ ir\sin\theta \pm (d^2-r^2\sin^2\theta)^{1/2}\,] \,\bigr)^{1/2}
&\quad \text{if} \,\,\,\,
    d^2-r^2\sin^2\theta <0 \quad  (\text{broken symmetry}) 
,
\end{split}
\end{align}
are \emph{non-orthogonal} with respect to the conventional Hermitian inner product.
With their Hermitian adjoint states
\begin{equation}
\label{s4e9}
    \bra{\varphi_\pm} = \, \bigl( - i r \, \sin\theta \,\pm \,
    \mathrm{sgn}(d^2 - r^2 \sin^2\theta) (d^2-r^2\sin^2\theta)^{1/2} ,\, d \bigr)
    \bigr/ \mathcal{N}_\pm  
    \, ,
\end{equation}
the overlap between the energy eigenstates takes the form
\begin{align}
\label{s4e10-1}
\braket{\varphi_\pm \vert \varphi_\mp} = 
\mfrac{i r \sin\theta}{d^2} \,
[-i r \sin\theta \pm (d^2- r^2\sin^2\theta)^{1/2} \, ] 
\quad &\text{and} \quad
\braket{\varphi_\pm \vert \varphi_\pm} = 1 \quad \text{if} \,\,\,\,
d^2-r^2\sin^2\theta >0 \quad  (\text{unbroken symmetry}) \\
\label{s4e10-2}
\braket{\varphi_\pm \vert \varphi_\mp} = 
\bigl\lvert \, \mfrac{d}{r \, \sin\theta } \, \bigr\rvert
\quad &\text{and} \quad
\braket{\varphi_\pm \vert \varphi_\pm} = 1 \quad \text{if} \,\,\,\, 
d^2-r^2\sin^2\theta <0 \quad  (\text{broken symmetry}) 
\, .
\end{align}
\end{widetext}
Note that the transition amplitudes $\braket{\varphi_\pm \vert \varphi_\mp}$ do not vanish generally and orthogonal eigenstates are only recovered in the Hermitian isolated-system limit $\theta \rightarrow 0 \lor\pi$, compare (\ref{s4e10-1}), or for a vanishing direct coupling $d$, see (\ref{s4e10-2}). 
Only in these limiting cases are the stationary eigenstate solutions represented by a set of antipodal points on the Bloch sphere within the Hermitian-adjoint formalism.

\section{Biorthogonal formalism}
\label{s5}

The breakdown of the orthogonality of eigenstates for the non-Hermitian Hamiltonian (\ref{s2e1}) in the Hermitian-adjoint formalism originates as a direct consequence of the absent symmetry of $\cH$ under Hermitian conjugation, $\cH \neq \cH^\dagger$, while simultaneously prescribing the dynamics of the adjoint (left-hand) quantum states (\ref{s4e1}) to be governed by the conjugate Hamiltonian $\cH^\dagger$.  
An orthogonal set of states is na\"ively constructed when the adjoint states instead obey the equation of motion
\begin{equation}
\label{s5e1}
    -i \partial_t \llangle \varphi \vert =  \llangle{\varphi} \vert \, \cH 
\, ,
\end{equation}
where we denote $\llangle \varphi \vert$ to indicate that this state is generally not the Hermitian adjoint of a state $\ket{\varphi}$ obeying (\ref{s2e4}).
The transition amplitude of the energy eigenstates then vanishes by construction,
\begin{equation}
\label{s5e2}
    (E_\pm-E_\mp)\llangle \varphi_\mp \ket{ \varphi_\pm} 
    = \llangle \varphi_\mp \vert \cH -\cH \ket{\varphi_\pm} = 0
\, ,
\end{equation}
instead of being affected by the anti-Hermitian contribution 
$\cH- \cH^\dagger = 2\cH_{ah}$, which generally prevents a vanishing overlap.
Together with the eigenstates $\ket{\varphi}$,
these adjoint states $\llangle \varphi \vert$ thus span a biorthogonal Hilbert space.

For the two-state system with balanced gain and loss, the rationale of (\ref{s5e1}) is founded in the symmetry of the Hamiltonian under combined parity reflection and time reversal,
\begin{equation}
\label{s5e3}
    \cH = \cH^{\cPT} \!\! , 
\,\,\text{where} \,\,
    \cH^{\cPT} \!\! = (\cPT)^{-1} \cH (\cPT) 
    \! = \cP^{-1} \cH^\dagger  \cP 
.
\end{equation}
Thus, the prescription of the adjoint states is based on a physical symmetry of the system instead of Hermitian adjointness.
In general terms, this treats the model as a \emph{pseudo}- (or \emph{quasi}-) Hermitian system, which satisfies the intertwining relation
\begin{equation}
\label{s5e3}
    \hat{g}\, \cH = \cH^\dagger\, \hat{g}  
\, .
\end{equation}
It endows the Hilbert space with a non-trivial metric $\hat{g}$, differing from
the conventional Euclidean metric of Hermitian systems, see e.g. \cite{SW-1972, SGH-1992, GHS-2008, B-2014} for detailed discussions. 
We remark that the choice of $\hat{g}$ is in general not unique, see also Sec.~\ref{s6} \cite{Jones2005, CMB2009, CMB2013}.
With respect to this state space, the non-orthogonality of eigenstates is lifted when the corresponding $\hat{g}$-inner product 
\begin{equation}
\label{s5e4}
    \bm{(} \cdot\vert\cdot \bm{)}_{\hat{g}} = 
    \braket{\cdot\vert \hat{g}\vert \, \cdot }
\end{equation}
is used. 
The system then behaves like familiar Hermitian models -- at least in the unbroken symmetry regime, discussed in the following subsection; 
the case of spontaneously broken $\cPT$ symmetry and its relation to (\ref{s5e1}) is addressed thereafter.

\subsection*{Biorthogonal-adjoint states (unbroken symmetry)}

In the case of the unbroken symmetry regime, where $\cH$ and $\cPT$ have simultaneous eigenstates, 
$\cPT \ket{\varphi} \propto \ket{\varphi}$, the $\cPT$-inner product has the form
\begin{equation}
\label{s5e5}
    \bm{(} \cdot \vert \cdot \bm{)}_{\cPT} = \braket{\cdot \vert \cP^T \vert \cdot}
\,.
\end{equation}
Here $\cP^T$ denotes the transpose; time reversal is accounted for through the use of the Hermitian-adjoint states in the Dirac notation on the right-hand side of equation (\ref{s5e5}).
The eigenstates now satisfy
\vspace{0.1cm}
\begin{equation}
\label{s5e6}
\begin{aligned}
    &\bm{(} \phi_\pm \vert \phi_\mp \bm{)}_{\cPT}= 0 \\[4pt]
    &\bm{(} \phi_\pm \vert \phi_\pm \bm{)}_{\cPT} = \pm 1
\end{aligned}
\qquad \text{if} \,\,\,\,
    d^2-r^2\sin^2\theta >0
.
\vspace{0.1cm}
\end{equation}
Here $\ket{\phi_\pm}$ correspond to the states (\ref{s4e7}) normalized to the $\cPT$-inner product by
$\mathcal{M} = [2d \,(d^2-r^2\sin^2\theta)^{1/2\,} ]^{1/2}$  instead of $\mathcal{N}_\pm$:
\begin{align}
\label{s5e6-2}
\begin{split}
	\ket{\phi_\pm} &= \, \bigl( i r \, \sin\theta\, 
    \pm (d^2-r^2\sin^2\theta)^{1/2} , d \bigr)^T
    \bigr/ \mathcal{M}  , \\
    \bra{\phi_\pm} &=  \bigl( - i r  \sin\theta \pm
    \mathrm{sgn}(d^2 \!- \! r^2 \sin^2\theta) (d^2 \!-\! r^2\sin^2\theta)^{1/2} , d \bigr)
    \bigr/ \mathcal{M}^* 
\, .
\end{split}
\end{align}
Evidently (\ref{s5e6}), and therefore (\ref{s5e5}), is appropriately orthogonal, but not positive definite.
A desired probabilistic interpretation of the inner product does, however, presuppose positive definiteness in addition to the orthogonality; 
here parity reflection and time reversal alone do not suffice.

Negative-norm states are, however, readily mended through the introduction of an operator \cite{CMB2002},
\begin{align}
\label{s5e7}
    \cC =& \,\cPT \ket{\phi_+}\!\bra{\phi_+}  + \cPT \ket{\phi_-}\!\bra{\phi_-} 
\\[5pt]
\label{s5e7-2}
    =& \, \mfrac{1}{\sqrt{d^2-r^2\sin^2\theta}}\begin{pmatrix}
        i r \sin\theta & d \\
        d & -i r \sin\theta
    \end{pmatrix}
,
\end{align}
labelled in analogy to the charge conjugation operator that resolves the similar issue of negative-probability states in fermionic quantum mechanics \cite{D-1942},
and which satisfies
\begin{equation}
\label{s5e8}
    \cC \ket{\phi_\pm} = \pm \ket{\phi_\pm} ,\quad\,
    \cC^2=\mathbbm{1} \, ,\quad \,
    [\cH, \cC]\! = [\cPT\!, \cC]\! = 0 
\, .
\end{equation}
Thus, in the state space with the non-trivial metric 
\begin{equation}
\label{s5e9}
    \hat{g} = (\cC\cP )^T
    =\, \mfrac{1}{\sqrt{d^2-r^2\sin^2\theta}}\begin{pmatrix}
        d &  -i r \sin\theta &\\
        i r \sin\theta & d
    \end{pmatrix} 
,
\end{equation}
and with the corresponding $\cCPT$-inner product
\begin{equation}
\label{s5e10}
    \bm{(} \cdot\, \vert \,\cdot \bm{)}_{\cCPT} = \braket{\cdot \,\vert (\cC\cP)^T \vert \,\cdot} ,
\end{equation}
the energy eigenstates of the system form a positive-definite orthogonal set, in line with a probabilistic interpretation.
In the Hermitian limit $\theta \rightarrow 0 \lor \pi$ the $\cC$ operator (\ref{s5e7-2}) takes the form of the parity operator $\cP$, so that the Hilbert space metric $\hat{g}$ in (\ref{s5e9}) trivializes to the conventional Euclidean form -- the $\cCPT$ symmetry of the Hamiltonian (\ref{s2e1}) reduces to its Hermiticity \cite{B-2005}. 

Notice that by identifying the adjoint states of the na\"ive approach (\ref{s5e1}) as 
\begin{equation}
\label{s5e11}
    \llangle \phi \vert = \bra{\phi} \,\hat{g} = \bra{\phi} ( \cC \cP)^T 
\end{equation}
we recognize a theory with a conventional Euclidean metric of the state space, but in which the dynamics of the adjoint eigenstates are governed by the $\cCPT$-conjugate Hamiltonian, instead of $\cH^\dagger$: 
\begin{equation}
\label{s5e12}
    -i \partial_t \, \llangle \phi \vert =  \llangle \phi \vert \, \cH^{\cCPT} =  \llangle \phi \vert \, \cH.
\end{equation}
This is also consistent with the definition of the eigenstates $\ket{\phi}$, which is to say that based on (\ref{s5e11}) and (\ref{s5e12}) the Hermitian adjoint states $\bra{\phi}$ indeed satisfy the Hermitian adjoint of the equation of motion (\ref{s2e4}).

The density matrix within the biorthogonal formalism
is now constructed using the adjoint states (\ref{s5e11}):
\begin{equation}
\label{s5e14}
	\rho_b = \sum_n p_n \ket{\phi_n}\!\llangle \phi_n\vert
\, ,  
\end{equation}
with $\sum_n p_n = 1$ initially.
From this, the evolution of the density is governed by the Liouville-von Neumann equation 
\begin{equation}
\label{s5e13}
    \partial_t \rho_b  = -i [\cH, \rho_b] = -i [\cH_h, \rho_b] + [\cH_{ah}, \rho_b] 
\end{equation}
and the time evolution of its trace vanishes,
\begin{align}
\label{s5e15}
    \partial_t \mathrm{Tr}[\rho_b] = 0
\, .
\end{align}
Notably, the description of the non-Hermitian system within this framework behaves identical to Hermitian systems in the conventional formalism and may be evaluated as such. 
In particular, we find that, analogous to closed-system dynamics, both the purity 
\begin{equation}
\label{s5e16}
\gamma_{b}(t) = \mathrm{Tr}[\rho_b^2(t)] ,
\qquad \partial_t \gamma_b(t) = 0,
\end{equation}
and the von Neumann entropy 
\begin{equation}
\label{s5e17}
s_{b}(t) = -\mathrm{Tr}[\rho_b(t) \ln(\rho_b(t) )] ,
\qquad \partial_t s_b(t) = 0,
\end{equation}
are time-independent within the biorthogonal formalism. 
 
We conclude that, unlike for the focus on the two-state subsystem in Sec.~\ref{s4},  
the environmental coupling mode is \emph{included} within the extent of the system described by the biorthogonal formalism perspective presented in this section, and emphasize that the density matrix $\rho_b$ in (\ref{s5e14}) differs from the Hermitian-adjoint density matrix $\rho$ in (\ref{s4e3}).

\subsection*{Biorthogonal-adjoint formalism (broken symmetry)}

Having mended the breakdown of the orthogonality of eigenstates for the non-Hermitian Hamiltonian in the unbroken $\cPT$-symmetry phase, 
an \emph{analytic continuation} of this perspective to $d^2-r^2\sin^2\theta <0 $ may, at first glance, appear to be a natural approach to describing the system in the broken regime.
This would, in particular, entail a continuation of the explicit form (\ref{s5e7-2}) of the $\cC$ operator in the unbroken phase, referred to hereafter as 
\begin{equation}
\label{s5e23}
\cC_u = \, \mfrac{1}{\sqrt{d^2-r^2\sin^2\theta}}\begin{pmatrix}
        i r \sin\theta & d \\
        d & -i r \sin\theta
    \end{pmatrix}
.
\end{equation}
However, when we evaluate the $\cC_u\cPT$-inner product of the biorthogonal formalism in the broken symmetry regime, we once more find the eigenstates to be non-orthogonal and, moreover, to have a vanishing norm,
\begin{equation}
\label{s5e24}
\begin{aligned}
&\bm{(} \phi_\pm \vert \phi_\mp \bm{)}_{\cC_u\cPT}= i \\
&\bm{(} \phi_\pm \vert \phi_\pm \bm{)}_{\cC_u\cPT} = 0
\end{aligned}
\qquad \text{if} \,\,\,\,
d^2-r^2\sin^2\theta <0,
\vspace{0.1cm}
\end{equation}
thus invalidating our initial incentive for the consideration of this formalism.

In fact, this is not a result of the analytic continuation of $\cC_u$, 
which, despite being anti-$\cPT$ symmetric in the broken regime, $\{\cC_u,\cPT \}=0$,
still satisfies  
\begin{equation}
\label{}
    \cC_u \ket{\phi_\pm} = \pm \ket{\phi_\pm} ,\quad\,
    \cC_u^2=\mathbbm{1} \, ,\quad \,
    [\cH, \cC_u]\! = 0 
\, ,
\end{equation}
so that notably $[\cH, \cC_u\cPT]=0$ still holds.
Rather, it is the spontaneous breaking of the $\cPT$ symmetry itself that causes this behavior.

In the broken $\cPT$ symmetry phase, while $[\cH, \cPT]=0$, the Hamiltonian $\cH$ and the symmetry operator $\cPT$ no longer have simultaneous eigenstates: $\cPT\ket{\phi} \not\propto \ket{\phi}$. 
Despite the symmetry of the Hamiltonian, the construction of a $\cPT$-inner product based on adjoint eigenstates governed by $\cH^{\cC_u\cPT}$ is therefore no longer equivalent to the (orthogonal) na\"ive approach of adjoint states governed by $\cH$:
\begin{equation}
\label{s5e25}
-i \partial_t \, \llangle \phi \vert =  \llangle \phi \vert \, \cH^{\cC_u\cPT} \neq  \llangle \phi \vert \, \cH.
\end{equation}
Since the energy eigenvalues of $\cH$ occur in complex conjugate pairs, one finds rather that the $\cPT$ eigenstates are proportional to the eigenstates of the complex conjugate eigenvalues:
$\cPT\ket{\phi} \propto \ket{\phi^*}$. Eigenstates corresponding to complex eigenvalues thus necessarily have a vanishing $\cPT$ norm \cite{M-2002}.
The effect of a charge-conjugation-like operator $\cC_u$, measuring the signature of the $\cPT$ norm, is thus inconsequential. 
This exemplifies a general result of pseudo-Hermitian theories: 
in a pseudo-Hermitian theory with metric $\hat{g}$ and non-real eigenvalues, which necessarily arise in complex-conjugate pairs, the $\hat{g}$-inner product is not orthogonal and the norm vanishes, cf.~\cite{M-2002}.

As a result, the biorthogonal-adjoint formalism based on the analytically continued operator $\cC_u$ fails to provide a satisfactory foundation for the system in the spontaneously broken $\cPT$-symmetry regime. 

Nonetheless, we observe that an alternative continuation of the $\cC$ operator 
based on its construction in (\ref{s5e7}), now using the eigenstates (\ref{s5e6-2})
with $\mathrm {sgn}(d^2-r^2\sin^2\theta)= -1$ for the broken symmetry phase,
yields 
\begin{widetext}
\vspace{0.1cm}
\begingroup
\setlength\arraycolsep{1.2pt}
\begin{equation}
\begin{split}
\label{s5e26}
    \cC_b = \cPT \ket{\phi_+}\!\!\bra{\phi_+}  +\! \cPT \ket{\phi_-}\!\!\bra{\phi_-}
    =  \mfrac{1}{\sqrt{r^2\sin^2\theta-d^2}}
    \begin{pmatrix}
        i r \sin\theta & d \\
        d & -i r \sin\theta
    \end{pmatrix}
    + \mfrac{2\sqrt{r^2\sin^2\theta -d^2}}{d}   \,
    \begin{pmatrix}
        0 & 0 \\
        1 & 0
    \end{pmatrix} 
    = -i C_u +
    \mfrac{\sqrt{r^2\sin^2\theta -d^2}}{d}   \,
    \sigma_-
,
\end{split}
\end{equation}
\endgroup
\end{widetext}
where $\sigma_- = \frac{1}{2}(\sigma_1 - i \sigma_2)$ in terms of the Pauli matrices is the 
decoupled-theory creation operator: $\sigma_- \ket{0} = \ket{1}$.
This approach, instead of an analytic continuation of the explicit structure in (\ref{s5e7-2}),
preserves a positive-definite orthogonal $\cC_b \cPT$-inner product: \vspace*{-0.1cm}
\begin{equation}
\label{s5e27}
    \begin{aligned}
        &\bm{(} \phi_\pm \vert \phi_\mp \bm{)}_{\cC_b\cPT}= 0 \\
        &\bm{(} \phi_\pm \vert \phi_\pm \bm{)}_{\cC_b\cPT} = 1
    \end{aligned}
\qquad \text{if} \,\,\,\,
    d^2-r^2\sin^2\theta <0
. \vspace*{-0.1cm}
\end{equation}

Notably, $\cC_b$ is an involution, $\cC_b^2 = \mathbbm{1}$, just like $\cC_u$.
But due to the $\sigma_-$ term, $\cC_b$ is neither $\cPT$ nor anti-$\cPT$ symmetric, because while $[i\cC_u, \cPT]=0$ in the broken regime, \vspace{-0.6cm}
\begin{equation}
\label{s5e29}
   (\cPT)^{-1}\,  \cC_b\,  (\cPT) = -i \cC_u + 2\delta \sigma_+ \neq \cC_b
,\vspace*{-0.1cm}
\end{equation}
where $\sigma_+ = \frac{1}{2}(\sigma_1 + i \sigma_2)$ 
is the decoupled-theory annihilation operator.
Moreover, the Hamiltonian $\cH$ is not $\cC_b$ symmetric, 
\vspace*{-0.1cm}
\begin{equation}
\label{}
   (\cC_b)^{-1}\, \cH \, (\cC_b ) =     \cC_b \, \cH \, \cC_b = 2r \cos\theta  -\cH
   \vspace*{-0.1cm}
\end{equation}
and thus not $\cC_b\cPT$ symmetric either, $[\cC_b\cPT, \cH] \neq 0$.

Despite (\ref{s5e29}), the correspondence between $\cC_b\cPT$ symmetry and pseudo-Hermiticity with respect to the metric $\hat{g}= (\cC_b\cP)^T$ remains intact, so that while the eigenstates of the theory are orthogonal in the $\cC_b \cP\cT$-inner product, the Hamiltonian is not pseudo-Hermitian with regard to this metric.
Once again, this realizes a general underlying result \cite{M-2002}: 
a theory with orthogonal $\hat{g}$-inner product and non-real eigenvalues is not $\hat{g}$-pseudo-Hermitian.

Nonetheless, this in itself is not at odds with our expectation of a system with an overall gain -- 
the presence of a generative environmental coupling mode, whether rooted in an external environment as in Secs.~\ref{s3} and \ref{s4}, 
or included within the extent of the system as a source, as here in Sec.~\ref{s5}, would be at odds with Hermiticity or probabilistic pseudo-Hermiticity. 

\begin{figure*}[t]
\centering
\includegraphics[width=0.95\textwidth]
{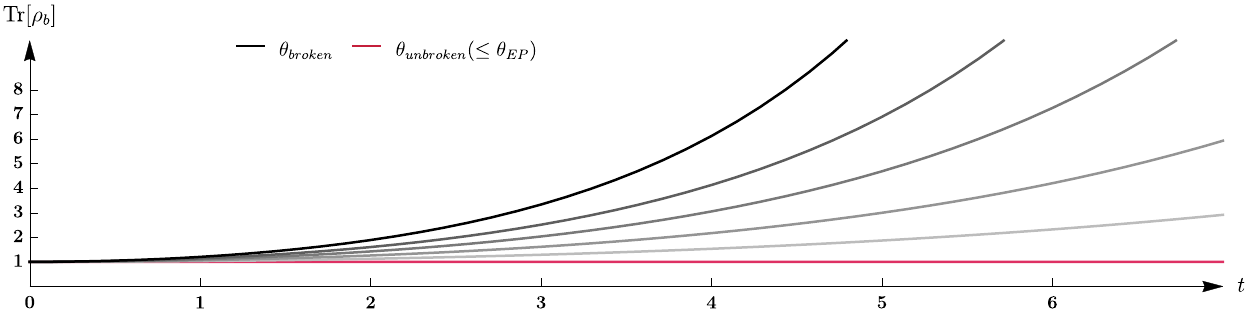}
\caption{
\label{f5}
Time evolution of $\mathrm{Tr}[\rho_b]$. Illustrative example with $r=1$, $d= 0.95$ and angles $\theta \in [0,\pi/2]$; the exceptional point lies at $\theta_{EP} \approx 1.253 $ (red). (Initial state  
$\ket{0}\!\bra{0} / [  \bm{(} 0 \vert 0  \bm{)}_{\cCPT}]^2$, normalized within the $\cCPT$-inner product.)
}
\end{figure*}
Since the $\cC_b\cPT$ symmetry of $\cH$ is broken, the positive definite and orthogonal 
$\cC_b\cPT$-inner product can not be time-independent,
\begin{align}
\label{s5e30}
\begin{split}
    &i \partial_t \bm{(} \psi \vert \phi \bm{)}_{\cC_b\cPT}
    = \,i \partial_t \bra{\psi} (\cC_b \cP)^T \ket{\phi} \\
    &= \bra{\psi}  (\cC_b \cP)^T \cH - (\cC_b \cP)^T \cH^{\cC_b\cPT} \ket{\phi}  \neq 0
.
\end{split}
\end{align}
Similarly, 
using that $\cH-\cH^{\cC_b\cPT} = 2\,(d^2-r^2\sin^2\theta)^{1/2} \, \cC_u$,
the density matrix $\rho_b$ (\ref{s5e14}) now satisfies the Liouville-von Neumann equation of the form \vspace*{-0.1cm}
\begin{align}
\label{s5e31}
\begin{split}
    \partial_t \rho_b =& -i (\cH \,\rho_b - \rho_b \, \cH^{\cC_b\cPT} ) \\
    =& -i [\cH, \rho_b] - 2  (r^2\sin^2\theta-d^2)^{1/2}   \, \rho_b \, \cC_u  
\, . \vspace*{-0.1cm}
\end{split}
\end{align}

\begin{figure*}[t]
\centering
\subfloat[\centering purity]{
\includegraphics[width=0.47\textwidth]
{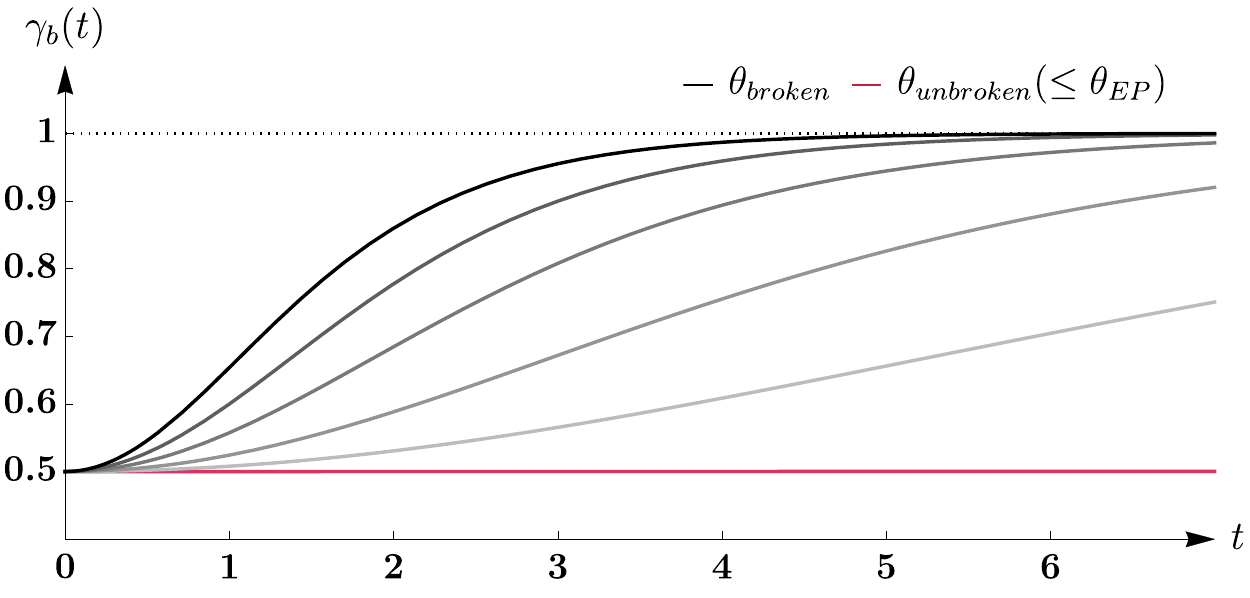}
\label{f6a}
}
\subfloat[\centering entropy]{
\includegraphics[width=0.47\textwidth]
{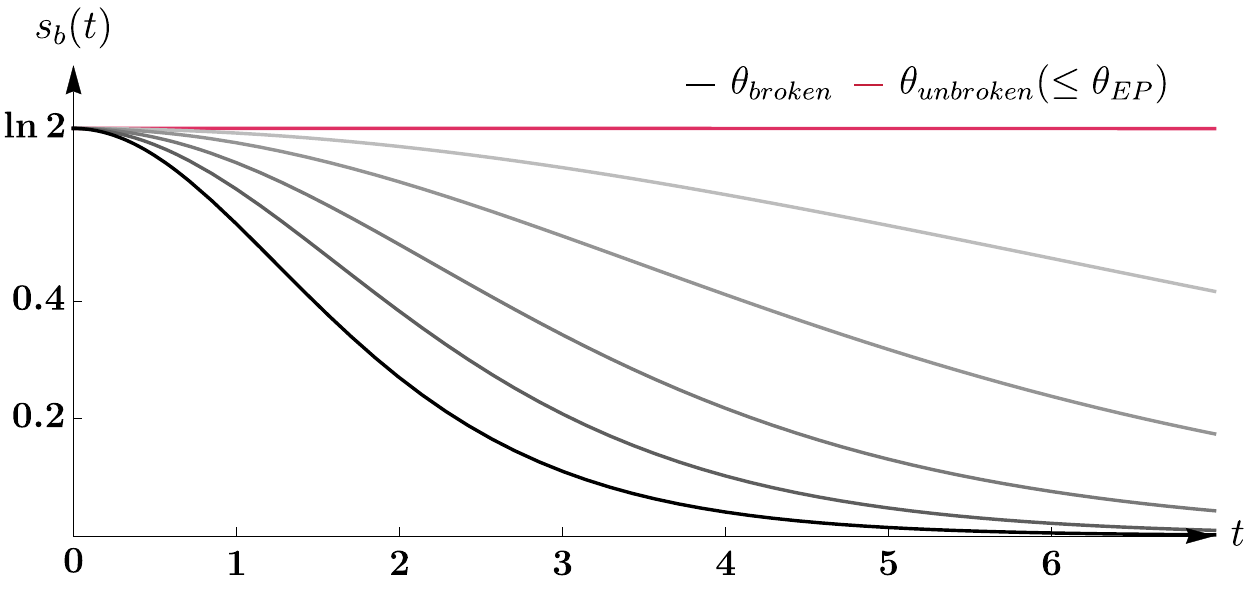}
\label{f6b}
}
\caption{
\label{f6}
Time evolution of the purity $\gamma_b(t)$ and the entropy $s_{b}(t)$. Illustrative example with $r=1$, $d= 0.95$ and angles $\theta \in [0,\pi/2]$; the exceptional point lies at $\theta_{EP} \approx 1.253 $ (red). (Maximally mixed initial state within the biorthogonal formalism with $\cCPT$-inner product.)
}
\end{figure*}
\noindent
Thus, the time-evolution of its trace within the $\cC_b\cPT$-inner product is not conserved
\vspace*{-0.2cm}
\begin{equation}
\label{s5e32}
    \partial_t \mathrm{Tr}[\rho_b] = -2 (r^2\sin^2\theta-d^2)^{1/2}\,\mathrm{Tr}[ \cC_u\, \rho ] 
\, . \vspace*{-0.1cm}
\end{equation}
It describes open-system dynamics akin to the behavior seen in the Hermitian-adjoint formalism, compare (\ref{s4e4}), except that here the underlying inner product is positive definite and orthogonal, and thus admits a probabilistic interpretation.

The evolution of the trace of the density matrix in the biorthogonal-adjoint formalism is illustrated in Fig.~\ref{f5}.
Throughout the unbroken symmetry regime, where the system admits a probabilistic pseudo-Hermitian description,  $\mathrm{Tr}[\rho_b]$ remains static (shown in red). 
The oscillatory behavior of the Hermitian-adjoint formalism,  which was associated with the subsystem perspective, is no longer present in the biorthogonal-adjoint formalism, which internalizes the environmental coupling mode.   
In the broken symmetry regime, however, an asymptotic growth of $\mathrm{Tr}[\rho_b]$ is found. 
This is qualitatively similar to the growth within the Hermitian-adjoint formalism, cf. Fig. \ref{f3}, which was sourced from a generative environmental coupling mode; here in the biorthogonal-adjoint formalism, this generative mode is internalized and acts as a source, that feeds the growth dynamics.      

Remarkably, the $\cPT$-charge-like operator $\cC_u$, constructed in the unbroken symmetry regime, arises in the spontaneously broken $\cPT$-symmetry phase as a 
source term driving the evolution (\ref{s5e31}) of the density $\rho_b$.
Rewriting (\ref{s5e32}) as 
\vspace*{-0.2cm}
\begin{equation}
\label{}
    \partial_t \rho_b = -i [(\cH+ \mfrac{i}{2} \Gamma\, \cC_u ), \rho_b] - \mfrac{1}{2} \Gamma \, \{\cC_u, \rho_b\}  
\, ,
\vspace*{-0.1cm}
\end{equation}
with $\Gamma = 2 (r^2\sin^2\theta-d^2)^{1/2}$,
further highlights a connection to established open-system descriptions in the form of the 
Lindblad master equation \cite{L-1976} without quantum-jump terms (semi-classical limit) \cite{ MMCN2019,RPCB2022, OZ2021}. 
The parameter $\Gamma$, quantifying the distance to the exceptional point, acts as a damping rate of the dissipator. 
\pagebreak

Paralleling the discussion in Sec.~\ref{s4} we can furthermore study the behavior of the purity 
\vspace*{-0.2cm}
\begin{equation}
\label{s5e33}
\gamma_b(t) = \mathrm{Tr}[\tilde{\rho_b}^2(t)] 
\vspace*{-0.2cm}
\end{equation}
and the von Neumann entropy
\vspace*{-0.1cm}
\begin{equation}
\label{s5e34}
s_{b}(t) = -\mathrm{Tr}[\tilde{\rho_b}(t) \ln( \tilde{\rho_b}(t) )] ,
\vspace*{-0.1cm}
\end{equation}
expressed in terms of the normalized density matrix $\tilde{\rho_b}(t) = \rho_b(t)/\mathrm{tr}[\rho_b(t)]$, which extend their 
previous definitions in (\ref{s5e16}) and (\ref{s5e17}) into the broken symmetry regime.
Their behavior is exemplified in Fig.~\ref{f6} for a maximally mixed initial state in the biorthogonal-adjoint formalism,
$\rho_{b}(0) = \tfrac{1}{2} (\ket{\phi_+}\! \llangle \phi_+\vert + \ket{\phi_-}\! \llangle \phi_-\vert )$.
In the unbroken symmetry regime (red), we find the expected closed-system evolution of a static minimal purity $\gamma_{b}= \tfrac{1}{2}$  and a corresponding static maximal entropy $s_{b}= \ln 2$. Upon transitioning into the regime of spontaneously broken $\cPT$ symmetry, an asymptotic purification of the initial state ($\lim_{t\to \infty}\gamma_b(t) = 1$) with a corresponding asymptotically vanishing entropy is found instead, mirroring the dynamics obtained within the Hermitian-adjoint formalism.

\section{Isospectral mapping} 
\label{s6}

The perspective taken in the biorthogonal formalism can be differentiated from the Hermitian-adjoint perspective further by considering the non-trivial metric (\ref{s5e9}) through the lens of a similarity transformation 
\vspace*{-0.1cm}
\begin{equation}
\label{s5e19}
    \eta =\sqrt{(\cC\cP)^T}, 
    \qquad \eta^\dagger \eta 
    = \eta^2 
    = \hat{g}.
\vspace*{-0.1cm}
\end{equation}
This maps the Hamiltonian $\cH$ to a spectrally equivalent model with the Hamiltonian 
\vspace*{-0.1cm}
\begin{equation}
\label{s5e20}
    h = \eta \, \cH \,\eta^{-1} 
    \vspace*{-0.1cm}
\end{equation}
and with a conventional flat state space, spanned by the orthonormal eigenstates $\eta \ket{\phi}$ and $\llangle \phi \vert \eta^{-1} = \bra{\phi} \eta^\dagger$, governed by the equations of motion
\begin{align}
\label{}
\begin{split}
    i \partial_t [\eta \ket{\phi}] = h \, [\eta \ket{\phi}] , \quad
    -i \partial_t [\bra{\phi} \eta^\dagger] = [\bra{\phi} \eta^\dagger] \, h
.
\end{split}
\end{align}

This construction is sometimes referred to as a Dyson map \cite{M-2010}, after an early example of a transformation taking fermionic Hermitian Hamiltonians without orthogonal eigenstates to equivalent pseudo-Hermitian bosonic Hamiltonians with an orthogonal basis \cite{D-1956}.

We comment that the similarity transformation $\eta$ inherits the non-uniqueness of the metric $\hat{g}$ through the ambiguity in choosing the $\cC$ operator, see for example \cite{Jones2005, CMB2009, CMB2013, Zno2006}.

\subsection*{Unbroken symmetry regime}

\vspace*{-0.2cm}
With the $\cC$ operator (\ref{s5e7}) (that is $\cC_u$) and the parity reflection operator $\cP$ in (\ref{s2e2})
one possible similarity transformation is found to be
\vspace*{-0.3cm}
\begin{widetext}
\begin{equation}
\label{s5e21}
    \eta =
    \mfrac{1}{2\sqrt[4]{d^2-r^2\sin^2\theta}}
    \begin{pmatrix}
       \sqrt{d+r\sin\theta} + \sqrt{d-r\sin\theta} & 
       -i (\sqrt{d+r\sin\theta} - \sqrt{d-r\sin\theta})\\
       i(\sqrt{d+r\sin\theta} - \sqrt{d-r\sin\theta})  & 
       \sqrt{d+r\sin\theta} + \sqrt{d-r\sin\theta}
    \end{pmatrix}
    , \quad \eta = \eta^\dagger
    , \quad \eta^{-1} = \eta^*
\end{equation}
\end{widetext}
and the Hamiltonian $h$, that is spectrally equivalent to $\cH$ in (\ref{s2e1}), is found to have the form \cite{OZ2020} 
\begin{equation}
\label{s5e22}
    h = 
    \begin{pmatrix}
        r \cos\theta & \sqrt{d^2-r^2 \sin^2\theta} \\
        \sqrt{d^2-r^2 \sin^2\theta} & r\cos\theta
    \end{pmatrix}
\, , \qquad h = h^\dagger 
.
\end{equation}
Due to the $\cC_u\cPT$ symmetry of the Hamiltonian $\cH$, the isospectral Hamiltonian $h$ is Hermitian: 
\begin{equation}
\label{}
    h^\dagger = (\eta^{-1})^\dagger \,\cH^\dagger\, \eta^\dagger = \eta \, [\hat{g}^{-1} \cH^\dagger\, \hat{g}] \, \eta^{-1} 
    = \eta \, \cH \, \eta^{-1} = h
.
\end{equation}

We emphasize that, despite this Hermiticity, the isospectrally mapped description differs in its perspective from the Hermitian-adjoint formalism outlined in Secs.~\ref{s3} and \ref{s4}, and rather reexpresses the biorthogonal-formalism perspective of Sec.~\ref{s5}: 
The mapped system (\ref{s5e22}) evidently corresponds to a two-state model without gain or loss terms that explicitly combines both the direct and the environmental coupling modes of $\cH$ in an (internal) effective direct coupling $(d^2-r^2 \sin^2 \theta)^{1/2}$.
In this, the isospectral mapping explicates the perspective of the \emph{biorthogonal-formalism description} 
in contrast to the Hermitian-adjoint formalism perspective, which describes subsystem dynamics in the presence of a distinct \emph{external} environmental coupling mode.
The biorthogonal and isospectral mathematical perspectives thus provide an equivalent physical description.

This further becomes apparent in the corresponding density matrix
\begin{equation}
\label{}
	\rho_\eta = \sum_n p_n \, \eta \ket{\phi_n}\!\bra{\phi_n} \eta^\dagger = \eta \, \rho_b \,\eta^{-1}
\, ,
\end{equation}
with $\sum_n p_n = 1$ initially.
Making use of the cyclic property of the trace, it immediately illustrates the identical behavior of the trace of the density matrix and the R\'enyi entropy to their respective biorthogonal-formalism counterparts, including the special cases of the purity and (using the replica trick) the von Neumann entropy.

\subsection*{Broken symmetry regime}

Lastly, we may also consider the system in the phase of spontaneously broken $\cPT$ symmetry from the perspective of an isospectral Hamiltonian $h_b$, obtained through a similarity transformation $\eta_b \!=\! \sqrt{(\cC_b\cP)^T}$ based on $\cC_b$ (\ref{s5e26}):
\begin{widetext}
\begin{align}
\label{s5e35}
&\eta_b = 
    \mfrac{1}{2 \sqrt[4]{-c^2}}
    \begin{pmatrix}
    \omega_{+}  \!+\! \frac{c^2}{\sqrt{c^4+c^2}} \,\omega_{-} &
    \frac{-is}{\sqrt{c^4+s^2}} \,\omega_{-} \\
    \frac{is}{\sqrt{c^4+s^2}}  \,\omega_{-} &
    \omega_{+} \!-\! \frac{c^2}{\sqrt{c^4+c^2}} \, \omega_{-}
    \end{pmatrix}    
, \quad \, \eta_b = \eta_b^\dagger, \quad \,
\eta_b^ {-1} = 
    \mfrac{1}{2 \sqrt[4]{-c^2}}
    \begin{pmatrix}
    \omega_{+}  \!-\! \frac{c^2}{\sqrt{c^4+c^2}} \,\omega_{-} &
    \frac{is}{\sqrt{c^4+s^2}} \,\omega_{-} \\
    \frac{-is}{\sqrt{c^4+s^2}}  \,\omega_{-} &
    \omega_{+} \!+\! \frac{c^2}{\sqrt{c^4+c^2}} \,\omega_{-}
    \end{pmatrix}  
,
\end{align}
\vspace{0.1cm}
where $s = r\sin\theta/d$, $c = \sqrt{d^2-r^2\sin^2\theta}/d$,  and $\omega_{\pm} = \sqrt{s^2+\sqrt{c^4+s^2}} \pm \sqrt{s^2-\sqrt{c^4+s^2}}$.
The resulting isospectral Hamiltonian has the form
\begin{equation}
\label{s5e35}
h_b = 
    \begin{pmatrix}
    r\cos\theta + i r\sin\theta\, (\frac{c^4+\sqrt{-c^2}}{c^4+s^2}) &
    \sqrt{r^2\sin^2\theta-d^2}\, (\frac{c^2+s^2\sqrt{-c^2}}{c^4+s^2}) \\[5pt]
    -\sqrt{r^2\sin^2\theta-d^2}\, (\frac{c^2+s^2\sqrt{-c^2}}{c^4+s^2}) &
    r\cos\theta - i r\sin\theta \,(\frac{c^4+\sqrt{-c^2}}{c^4+s^2})
    \end{pmatrix}   
.
\end{equation}
\end{widetext}
Notably this model is generally not Hermitian, because of the absent $\cC_b\cPT$ symmetry of the Hamiltonian $\cH$:  
\begin{equation}
\label{s5e36}
    h^\dagger = (\eta^{-1})^\dagger \,\cH^\dagger\, \eta^\dagger = \eta \, [\hat{g}^{-1} \cH^\dagger\, \hat{g}] \, \eta^{-1} 
    \neq \eta \, \cH \, \eta^{-1} 
.
\vspace{-0.01cm}
\end{equation}

Nor is it $\cPT$ symmetric, due to the skew-symmetric direct  coupling terms. 
This directionality is in qualitative agreement with the expectation gained from the biorthogonal-formalism description of the broken symmetry regime: the directed dynamics are driven by the presence of a source within the system, arising from the internalized generative environmental coupling mode.


A special case of (\ref{s5e35}) is the limit of vanishing direct coupling $d$, in which, despite the non-Hermiticity of the model, the eigenstates of the Hermitian-adjoint formalism are orthogonal to begin with, cf.~(\ref{s4e10-2}). The isospectrally mapped Hamiltonian $h_b$ here coincides with the original Hamiltonian $\cH$ in (\ref{s2e1}).

\section{Conclusion}
\label{s7}

The entropy of equilibrated closed (Hermitian) systems remains constant over time. 
In this study we have investigated the entropy dynamics for the open two-state system with balanced gain and loss -- an exemplary case of $\cPT$-symmetric models, which are an intermediary between closed and open, Hermitian and non-Hermitian, systems. 

Three commonly-used approaches were contrasted and their perspectives on the system differentiated: 
(a) applying the standard Hermitian framework to the non-Hermitian system, which describes a subsystem perspective of the two states of the model coupled to an external environment;
(b) the biorthogonal framework, which internalizes the environmental coupling as an additional coupling mode into the scope of the system and gives rise to a state space with non-trivial metric;
and (c) the isospectral mapping approach, which utilizes a similarity transformation to render the state space Euclidean again, while keeping with the biorthogonal perspective of an internalized environmental coupling mode. 

We have first examined the phase portrait of the model in the Hermitian formalism (a) on the Bloch sphere, elucidating the change in the behavior of the state dynamics as the spontaneously $\cPT$ symmetry breaking phase transition is crossed. This offers an illustrative representation on how the environmental coupling mode impacts the subsystem dynamics. It showcases in particular a source and sink flow between the stationary states in the broken symmetry regime, which clearly demonstrates the effect of the dominant unidirectional environmental coupling mode on the overall dynamics in this phase.

In the regime of unbroken $\cPT$ symmetry, where the direct coupling of the two states supersedes the strength of the environmental coupling, a characteristic feature of the subsystem perspective 
(a)
is the presence of \emph{purity oscillations} and according \emph{entropy-revival dynamics}, that indicate a periodic evolution of the entanglement structure between the subsystem and the environment. In contrast, these oscillations are not apparent within the biorthogonal and isospectral-mapping approaches 
(b) and (c)
in which the environmental coupling is internalized. The constant entropy evolution in these formalisms is reflective of their description of the model to behave like a Hermitian - closed - system within a non-trivial state space.  

While the subsystem perspective may be easier to access for experimental realization of $\cPT$ systems \cite{WZZ2021,DMP2021, SRDCW-2019, KMCW2013, LHM2019}, theoretical discussions utilizing projection methods or perturbation techniques rely on an orthogonal inner-product structure of the underlying state space, which is (usually) missing in the standard Hermitian formalism when applied to non-Hermitian models. The restoration of this structure is the foundation of the biorthogonal approach.
But addressing the phase transition into the regime of spontaneously broken $\cPT$ symmetry is not obvious. 

We showed that an analytic continuation of the inner product structure based on the introduction of the $\cPT$ charge operator $\cC$ fails to provide a 
probabilistic (positive definite) orthogonal inner product in the broken regime, illustrating a general property of pseudo-Hermitian models. 
However, 
our introduction of
an \emph{operator-based} continuation 
$\cC_b$ that is
rooted in the construction of the $\cC$ operator successfully restores such a structure, but explicitly breaks the $\cCPT$ symmetry (and with it pseudo-Hermiticity) of the system.
This is in agreement with expectation, because in the broken symmetry phase the environmental coupling supersedes the direct coupling mode of the model and becomes \emph{generative} in the sense that it feeds an overall density growth of the system. It acts as an effective source. Thus, genuine non-Hermitian (open) system dynamics are to be expected. 
The introduction of the operator $\cC_b$ thus enables the investigation of the purity and entropy dynamics throughout the broken symmetry regime, allowing for a complete comparison of the system behavior within all three formalisms (a) to (c) throughout both the phase of spontaneously broken and preserved $\cPT$ symmetry.   
A prominent resulting feature is the \emph{asymptotic purification} of the system in the broken regime, with an according \emph{asymptotically vanishing entropy}, which is found irrespective of the perspective taken.  

The open-system nature of the broken symmetry regime becomes further apparent by rewriting the Liouville-von Neumann equation governing the dynamics of the biorthogonal density matrix in the form of a Lindblad master equation without quantum-jump terms. Remarkably, the $\cPT$ charge operator $\cC$ operator of the unbroken symmetry regime here arises as a source term of the evolution. 
This indicates a fundamental connection between a closed-system description of Hermitian systems in nontrivial Hilbert spaces (unbroken $\cPT$ symmetry regime) and established non-Hermitian open-system descriptions based on the master equation through an underlying spontaneous breakdown of the time-reversing reflection symmetry $\cPT$.

\bibliographystyle{alpha}

\end{document}